\documentclass[conference,a4paper]{IEEEtran}
\usepackage{mathptmx}
\usepackage{amsmath, amssymb, bm, color, graphicx, latexsym, amsthm}
\newtheorem{defi}{Definition}
\newtheorem{theorem}{Theorem}
\newcommand{\Eref}[1]{eq.~(\ref{#1})}
\newcommand{\Fref}[1]{Fig.~\ref{#1}}
\newcommand{\figwidth}{3.1in}
\newcommand{\entropy}{{\omega }}

\begin{document}

\sloppy

\title{Replica Symmetric Bound for \\
Restricted Isometry Constant}

\author{
  \IEEEauthorblockN{Ayaka Sakata}
  \IEEEauthorblockA{The Institute of Statistical Mathematics\\
  Tachikawa, Japan\\
  Theoretical Biology Laboratory\\
    RIKEN, Wako, Japan\\
    Email: ayaka@ism.ac.jp}
  \and
  \IEEEauthorblockN{Yoshiyuki Kabashima}
  \IEEEauthorblockA{Dep. of Computational Intelligence \& Systems Science\\
    Tokyo Institute of Technology\\
    Yokohama, Japan\\
    Email: kaba@dis.titech.ac.jp}
}



\maketitle

\begin{abstract}
We develop a method for
evaluating restricted isometry constants (RICs).
This evaluation is reduced to the identification of 
the zero-points of entropy density
which is defined for submatrices that are composed of
columns selected from a given measurement matrix.
Using the replica method developed 
in statistical mechanics,
we assess RICs for Gaussian random matrices under
the replica symmetric (RS) assumption.
In order to numerically validate the adequacy of our analysis,
we employ the exchange Monte Carlo (EMC) method,
which has been empirically demonstrated to achieve
much higher numerical accuracy than naive Monte Carlo methods.
The EMC method suggests that our theoretical estimation of an RIC
corresponds to an upper bound that is
tighter than in preceding studies.
Physical consideration indicates that
our assessment of the RIC could be improved 
by taking into account the
replica symmetry breaking.
\end{abstract}

\section{Introduction}

The 
signal processing paradigm of
compressed sensing (CS)
enables a substantially more effective sampling than that required by the conventional sampling theorem \cite{Nyquist}.
CS is applied to problems in various fields, in which the acquisition of data is quite costly,
such as astronomical and medical imaging \cite{Lustig,Starck}.
The CS performance is mathematically analyzed
using the problem settings of a randomized linear observation \cite{CRT,Donoho}. Here,
$\bm{A}\in\mathbb{R}^{M\times N}$ is the given observation matrix,
and CS endeavors to reconstruct the $S$-sparse signal
$\bm{x}\in\mathbb{R}^N$ that has $S(<N)$ nonzero components
from observation $\bm{y}=\bm{Ax}$.

A widely used strategy for the reconstruction of this signal
is the $\ell_1$ minimization, 
which corresponds to the relaxed problem of
$\ell_0$ minimization. 
A key quantity used to 
analyze the $\ell_0$ and $\ell_1$ minimization strategies
is the restricted isometry constant (RIC)
\cite{CT}.
Literally
evaluating an RIC
requires the computation of maximum and minimum eigenvalues of
$N!/(S!(N-S)!)$
submatrices that are
generated by extracting $S$-columns from $\bm{A}$,
which is computationally infeasible.
In the case of Gaussian random matrices of $\bm{A}$, 
the upper bound for the RIC is estimated
using the large deviation property
without direct computation of the eigenvalues
\cite{CT,BCT,BT}.

This paper 
proposes a theoretical scheme
for the direct estimation of 
the RICs.
In order to do this, we evaluate
the entropy density of the submatrices that 
provide
a given value of
the maximum/minimum eigenvalues. 
An RIC of matrix $\bm{A}$ is  
offered by the condition that the corresponding entropy vanishes.
Furthermore, in order to demonstrate our method's utility,
we apply our scheme to 
Gaussian random matrices,
using the replica method, and
compare the obtained result with that of earlier studies.

Our theoretical evaluation is also numerically assessed
using the exchange Monte Carlo (EMC) 
sampling \cite{EMC},
which is expected to achieve much higher numerical accuracy
than those of naive Monte Carlo schemes.
The EMC method enables effective sampling, avoiding
entrapment at local minima,
which limits the effectiveness of naive Monte Carlo sampling to capture
the true behavior \cite{DonohoTsaig}.
Numerical results suggest that
our scheme currently provides the tightest RIC upper bound,
which could be further tightened 
by taking into account the replica symmetry breaking (RSB).

\section{Restricted isometry constant}
In the following, we assume that 
$\forall{\bm{A}}\in \mathbb{R}^{M \times N}$ is 
normalized so as to (typically) satisfy $(\bm{A}^{\rm T}\bm{A})_{ii}=1$
for all $i\in\{1,\cdots,N\}$.
\begin{defi}[Restricted isometry constants]
\label{definition}
A matrix $\bm{A}\in\mathbb{R}^{M\times N}$
satisfies the restricted isometry property (RIP)
with RIC $0<\delta_S^{\rm
 min}\leq\delta_S^{\rm max}$
if
\begin{align}
(1-\delta_S^{\rm min})||\bm{x}||_F^2\leq ||\bm{Ax}||_F^2\leq
 (1+\delta_S^{\rm max})||\bm{x}||_F^2
\label{eq:def_RIC}
\end{align}
holds for any $S$-sparse vector
$\bm{x}\in\mathbb{R}^N$,
in which $S$ is the number of non-zero components.
\end{defi}
The original work presented by Cand\`{e}s et al. \cite{CRT}
addresses symmetric RIC $\delta_S=\max[\delta_S^{\rm
min},\delta_S^{\rm max}]$.
An RIC indicates how close the space, which is 
spanned by the $S$-columns of $\bm{A}$, is to an orthonormal system. If an RIC is small, the linear transformation
performed using $\bm{A}$ is nearly an orthogonal transformation.

The 
symmetric RIC provides sufficient conditions for
the reconstruction of $S$-sparse vector $\bm{x}$ in
underdetermined linear system $\bm{y}=\bm{Ax}$
using $\ell_0$ and $\ell_1$ minimization \cite{CT}.
\begin{theorem}
Let $\bm{A}\in\mathbb{R}^{M\times N}$ and $\bm{x}\in\mathbb{R}^N$
with $M<N$,
and consider the linear equation $\bm{y}=\bm{Ax}$.
If 
$\delta_{2S}<1$,
a unique $S$-sparse solution exists and is the sparsest solution
to $\ell_0$ problem
\begin{align}
\min_{\bm{x}}||\bm{x}||_0,\mbox{ subject to }\bm{y}=\bm{Ax}.
\end{align}
Also, if $\delta_{2S}<\sqrt{2}-1$,
the $S$-sparse solution to $\ell_1$ problem
\begin{align}
\min_{\bm{x}}||\bm{x}||_1,\mbox{ subject to }\bm{y}=\bm{Ax}
\end{align}
is uniquely identified as the sparsest solution and equals the $\ell_0$ problem's solution.
\label{theorem:RIP}
\end{theorem}
It should be noted that $\delta_S^{\rm min}$
and $\delta_S^{\rm max}$ do not
increase or decrease at the same rate, and
asymmetric RICs improve the condition of $\ell_1$
reconstruction \cite{FL}.
\begin{theorem}
Consider 
the same problem settings as in
Theorem \ref{theorem:RIP}.
If
$(4\sqrt{2}-3)\delta_{2S}^{\rm min}+\delta_{2S}^{\rm max}<4(\sqrt{2}-1)$,
then the unique $S$-sparse solution is the sparsest solution to the $\ell_1$ problem
and equals the solution to the $\ell_0$ problem \cite{FL}.
\label{theorem:RIP2}
\end{theorem}

RIC evaluation
is also a fundamental linear algebra problem \cite{BCT,BT} because
RICs clearly relate to the eigenvalues of
Gram matrices.
Let $T\subseteq V=\{1,\cdots,N\},~|T|=S$
be the position of the nonzero elements of
$S$-sparse vector $\bm{x}$.
The product $\bm{Ax}$ equals $\bm{A}_T\bm{x}_T$,
where $\bm{A}_T$ is the submatrix that consists of $i\in T$
columns of $\bm{A}$
and where $\bm{x}_T=\{x_i|i\in T\}$.
For any realization of $T$, the following holds.
\begin{align}
\nonumber
\lambda_{\rm min}(\bm{A}_T^{\rm T}\bm{A}_T)||\bm{x}_T||_F^2\leq||\bm{A}_T\bm{x}_T||_F^2\leq\lambda_{\rm max}(\bm{A}_T^{\rm T}\bm{A}_T)||\bm{x}_T||_F^2
\end{align}
Here, $\lambda_{\rm min}(\bm{B})$ and $\lambda_{\rm max}(\bm{B})$
denote the minimum and maximum eigenvalues of $\bm{B}$, respectively,
and superscript ${\rm T}$ denotes the matrix transpose.
Therefore, the following expression of the RIC is equivalent to
\Eref{eq:def_RIC}:
\begin{align}
\delta_S^{\rm min}=1-\lambda^*_{\rm min}(\bm{A};S),~~\delta_S^{\rm max}=\lambda^*_{\rm max}(\bm{A};S)-1,
\label{eq:def_RIC2}
\end{align}
in which
\begin{align}
 \lambda_{\rm min}^*(\bm{A};S)&=\min_{T:T\subseteq V,|T|=S}\lambda_{\rm
 min}(\bm{A}_T^{\rm T}\bm{A}_T), \label{eq:lambda_min1}\\
 \lambda_{\rm max}^*(\bm{A};S)&=\max_{T:T\subseteq V,|T|=S}\lambda_{\rm
 max}(\bm{A}_T^{\rm T}\bm{A}_T).\label{eq:lambda_max1} 
\end{align}
Literal evaluation of \Eref{eq:def_RIC2} requires
the calculations of the maximum and minimum eigenvalues of the 
$N!/\left (S!(N-S)! \right )$
Gram matrices $\{\bm{A}_T^{\rm T}\bm{A}_T\}$,
which is computationally difficult when $N$ and $S$ are large.
For typical Gaussian random matrices $\bm{A}$,
the RIC's upper bound is estimated
using large deviation properties of the maximum and minimum eigenvalues
of the Wishart matrix \cite{CT,BCT,BT}.

\section{Problem setup and formalism}

We estimate RICs in a different manner, and the following theorem is fundamental to our approach.
\begin{theorem}
Let $\bm{A}\in\mathbb{R}^{M\times N}$.
Then the 
minimum and maximum eigenvalues of $\bm{A}^{\rm T}\bm{A}$
are given by
\begin{align}
\lambda_{\min}(\bm{A}^{\rm
 T}\bm{A})&=-\lim_{\beta\to+\infty}\frac{2}{N\beta}\log Z(\bm{A};\beta),\label{eq:lambda_min2}\\
\lambda_{\max}(\bm{A}^{\rm
 T}\bm{A})&=-\lim_{\beta\to-\infty}\frac{2}{N\beta}\log Z(\bm{A};\beta),\label{eq:lambda_max2}
\end{align}
respectively, where
$Z(\bm{A};\beta)$ is defined using $\bm{u}\in\mathbb{R}^N$:
\begin{align}
Z(\bm{A};\beta)=\int d\bm{u}e^{\frac{\beta}{2}||\bm{Au}||_F^2}\delta(||\bm{u}||_F^2-N).
\end{align}
\label{theorem:lambda}
\end{theorem}
\noindent {\bf Proof:}
Applying identity $\delta(||\bm{u}||_F^2-N)\!=\!\beta/(4\pi)\times$
$\int d \eta \exp \left (-\beta \eta/2 (||\bm{u}||_F^2-N ) \right )$ gives us
\begin{align}
\nonumber
Z( \bm{A};\beta) \!= \!\frac{(2\pi)^{\frac{N}{2}\!-\! 1}}{2 \beta^{\frac{N}{2}\! - \!1}}
\!\! \int \!\!d\eta\exp\!\Big[\beta \Big\{\frac{N \eta}{2}\!-\!\frac{1}{2\beta }\!\sum_i\ln(\eta\!+\!\lambda_i) \Big\}\Big],
\end{align}
in which $\{\lambda_i\}$ is the $i$th eigenvector of $\bm{A}^{\rm T}\bm{A}$.
As $\beta\to +\infty$, the integral can be evaluated using the saddle point method,
which is dominated by $\eta=-\lambda_{\rm min}(\bm{A}^{\rm T}\bm{A}) + (N\beta)^{-1}+
o(\beta^{-1})$, where $o(\beta^{-1})$ represents the contribution from negligible terms compared
with $\beta^{-1}$.
This yields
\Eref{eq:lambda_min2}, and
eq.~(\ref{eq:lambda_max2}) is similarly obtained by applying
the saddle point method for $\beta \to -\infty$.
\hfill $\Box$

Theorem \ref{theorem:lambda} holds for all submatrices $\bm{A}_T$.
For mathematical convenience,
we introduce variables $\bm{c}\in\{0,1\}^N$
and define
\begin{align}
\nonumber
Z_c(\bm{c}, \bm{A};\beta)&=\!\!\int\!\! d\bm{u}P(\bm{u}|\bm{c})\exp\Big\{\!-\frac{\beta}{2}||\bm{A}(\bm{c}\circ\bm{u})||_F^2\Big\}\\
&\hspace{1.0cm}\times\delta(||\bm{c}\circ\bm{u}||_F^2-N),
\label{eq:Z_c_def}
\end{align}
where $\circ$ denotes the component-wise product,
and $P(\bm{u}|\bm{c})\propto\exp\Big(-\sum_{i=1}^N(1-c_i)u_i^2\slash 2\Big)$
is introduced in order to avoid the divergence
caused by integrating $u_i$ when $c_i=0$.
Let us define $\bm{c}(T) \in \{0,1\}^N$ to be
$(\bm{c}(T))_i=1$ for $i \in T$ and to be $(\bm{c}(T))_i=0$ otherwise.
The two functions $Z(\bm{A}_T;\beta)$ and $Z_c(\bm{c}(T),\bm{A};\beta)$
have a one-to-one correspondence:
$Z(\bm{A}_T;\beta)=Z_c(\bm{c}(T),\bm{A};\beta)$.
We write $\lambda_{\rm max}(\bm{c},\bm{A})$
and $\lambda_{\rm min}(\bm{c},\bm{A})$,
which are obtained by substituting $Z_c(\bm{c},\bm{A};\beta)$
into \Eref{eq:lambda_min2} and \Eref{eq:lambda_max2}, respectively.
Because $\lambda_{\rm max}(\bm{A}_T^{\rm
T}\bm{A}_T)=\lambda_{\rm max}(\bm{c}(T),\bm{A})$
and $\lambda_{\rm min}(\bm{A}_T^{\rm T}\bm{A}_T)=\lambda_{\rm min}(\bm{c}(T),\bm{A})$
naturally hold,
eqs.~(\ref{eq:lambda_min1}-\ref{eq:lambda_max1})
can be respectively rewritten as
\begin{align}
\lambda_{\rm min}^*(\bm{A};S)&=\min_{\bm{c}\in\bm{c}_S}\lambda_{\rm
 min}(\bm{c},\bm{A}),\label{eq:lambda_min3}\\
 \lambda_{\rm max}^*(\bm{A};S)&=\max_{\bm{c}\in\bm{c}_S}\lambda_{\rm
 max}(\bm{c},\bm{A}),\label{eq:lambda_max3}
\end{align}
where $\bm{c}_S$ denotes the set of configurations of $\bm{c}$
that satisfy $\sum_ic_i=S$. 

We hereafter focus on the situation in which both $M$ and $S$ are 
proportional to $N$ as
$M=N\alpha$ and $S=N\rho$, respectively, where $\alpha, \rho \sim O(1)$. 
Let us define the {\it energy} densities of $\bm{c}$ to be
$\Lambda_+(\bm{c}|\bm{A})\!=\!\lambda_{\rm min}(\bm{c},\bm{A})\slash 2$ and
$\Lambda_-(\bm{c}|\bm{A})\!=\!\lambda_{\rm max}(\bm{c},\bm{A})\slash 2$. 
Based on this, we introduce a {\em free entropy} density as
$\phi(\mu|\bm{A};\rho) = 
N^{-1}\log \! \left [ \!\sum_{\bm{c}}e^{-N\mu \Lambda_{{\rm sgn}(\mu)}(\bm{c}|\bm{A})}
\! \delta \! \left ( \!\sum_{i=1}^N \! c_i \!- \!N\rho \! \right ) \right ]$, where ${\rm sgn}(\mu)$ denotes the sign of $\mu$.
Eqs.~(\ref{eq:lambda_min2}-\ref{eq:lambda_max2}) offer its alternative expression 
\begin{align}
\phi(\mu|\bm{A};\rho)\!
=\!\!\!\lim_{\frac{\beta}{\mu} \to +\infty}\!\!\!\!\!N^{-1}\log \!
\Big[ \!\sum_{\bm{c}}\!Z_c^{\frac{\mu}{\beta}} \! (\bm{c},\bm{A};\beta)\delta \! \Big( \!\sum_{i=1}^Nc_i\!-\!N\rho\Big) \!\Big].
\label{free_entropy}
\end{align}


In addition, we represent
the number of $\bm{c}$
that correspond to $\Lambda_{\pm }(\bm{c}|\bm{A}) = \lambda\slash 2$
and satisfy $\sum_ic_i\!=\!N\rho $ as
$\exp \left (N \entropy_\pm(\lambda|\bm{A};\rho) \right )$
using {\em entropy} densities $\entropy_\pm(\lambda|\bm{A};\rho)$,
which are naturally assumed to be convex functions of $\lambda$.
Summation over the microscopic states of $\bm{c}$ is replaced
with the integral of $\lambda$ over the possible value of $\Lambda_\pm(\bm{c}|\bm{A})$:
\begin{align}
\nonumber
\phi(\mu|\bm{A};\rho)&=\frac{1}{N}\log\Big[\int \!\!d\lambda
 \exp\{-N\mu\lambda\!+\!N\entropy_{{\rm sgn(\mu)}}(\lambda|\bm{A};\rho)\}\Big]\\
&\to\max_{\lambda}\{-\mu\lambda+\entropy_{{\rm sgn}(\mu)}(\lambda|\bm{A};\rho)\},
\label{eq:phi_2}
\end{align}
in which the saddle point method is employed.
The maximizer of $\lambda$,
which corresponds to 
the typical energy value of $\bm{c}$ that is sampled following the weight  
$e^{-N\mu \Lambda_{{\rm sgn}(\mu)}(\bm{c}|\bm{A})}\!\delta\!(\sum_{i=1}^Nc_i-N\rho)$,
must satisfy
\begin{align}
-\mu+\frac{\partial\entropy_{{\rm sgn}(\mu)}(\lambda|\bm{A};\rho)}{\partial\lambda}=0.
\label{eq:lambda_mu}
\end{align}
Eq.~(\ref{eq:phi_2}) implies that $\phi(\mu|\bm{A};\rho)$
is obtained using the Legendre transformation of $\entropy_\pm(\lambda|\bm{A};\rho)$,
and the inverse Legendre transformation converts 
$\phi(\mu|\bm{A};S)$ to $\entropy_\pm(\lambda|\bm{A};\rho)$ as
\begin{align}
\entropy_{{\rm sgn}(\mu)}(\lambda|\bm{A};\rho)=\phi(\mu|\bm{A};\rho)-\mu\frac{\partial\phi(\mu|\bm{A};\rho)}{\partial\mu},
\end{align}
from the convexity assumption of $\entropy_\pm(\lambda|\bm{A};\rho)$.
A similar formalism has been introduced for investigating 
the geometrical structure of weight space in learning of multilayer neural networks \cite{Monasson1994}.

\begin{figure}
\begin{center}
\includegraphics[width=3.35in]{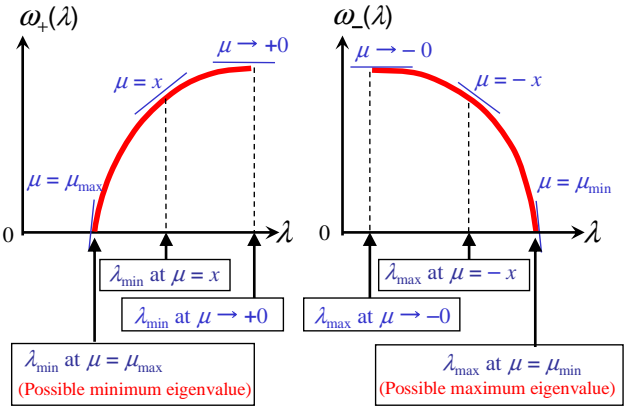}
\end{center}
\caption{Schematic picture of entropy curve and relationship to
 parameter $\mu$.}
\label{fig:complexity_illust}
\end{figure}

The relationships 
among $\mu$, $\lambda$, and $\entropy_\pm$
are illustrated in \Fref{fig:complexity_illust}.
Entropy densities $\entropy_+$ and $\entropy_-$ are convex increasing
and decreasing functions of $\lambda$, respectively.
According to \Eref{eq:lambda_mu},
the value of $\lambda$ at $\mu$ represents the point 
where the gradient of $\entropy_\pm$ equals $\mu$.
By definition, negative entropy values are not allowed, and
$\entropy_\pm(\lambda|\bm{A};\rho)<0$ implies that
no $\bm{c}$ simultaneously satisfies both $\Lambda_{\pm}(\bm{c}|\bm{A})=\lambda\slash
2$ and $\sum_ic_i=N\rho $.
Therefore, the $\lambda_\pm^*$ that produces 
$\entropy_\pm(\lambda_\pm^*|\!\bm{A};\rho)\!\!=\!\! 0$
is the 
possible minimum or maximum eigenvalue.
Hence, eqs.~(\ref{eq:lambda_min3}-\ref{eq:lambda_max3}) give us
\begin{align}
\lambda_{\rm min}^*(\bm{A};\rho)=\lambda_+^*,~~\lambda_{\max}^*(\bm{A};\rho)
=\lambda_-^*,
\end{align}
which are the typical values for $\mu=\mu_{\rm max}$ and $\mu=\mu_{\rm min}$,
respectively (\Fref{fig:complexity_illust}).

\section{RS analysis for Gaussian random matrix}

This section applies the methodology introduced in the previous section to
the case in which components of $\bm{A}$ 
are 
independently generated using a Gaussian distribution with mean 0 and variance 
$(N\alpha)^{-1}$.
In this case, $\phi(\mu|\bm{A};\rho)$ and $\entropy_\pm(\lambda|\bm{A};\rho)$
randomly fluctuate depending on $\bm{A}$.
However, for all $\epsilon>0$,
the probability that deviation from the typical values, 
$\phi(\mu;\rho)\equiv[\phi(\mu|\bm{A};\rho)]_A$
and $\entropy_\pm(\lambda;\rho)\equiv[\entropy_\pm(\lambda|\bm{A};\rho)]_A$,
is larger than $\epsilon$ tends to vanish as $N \to \infty$. Here,
$[\cdot]_A$ denotes the average of $\bm{A}$.
Therefore, typical properties can be characterized by evaluating
the typical values, $\phi(\mu;\rho)$ and $\entropy_\pm(\lambda)$, using
the replica method with the identity \cite{beyond,Nishimori}:
\begin{align}
[\log f(\bm{A})]_A=\lim_{n\to 0}
\frac{\partial}{\partial n} \log [f^n(\bm{A})]_A
\label{eq:def_replica}
\end{align}
where $f(\bm{A})$ is an arbitrary function.
When both $n$ and $m=\mu/\beta$ are positive integers, 
regarding 
$\sum_{\bm{c}}Z_c^{m}(\bm{c},\bm{A};\beta) \delta (\sum_{i=1}^Nc_i-N\rho)$
in \Eref{free_entropy}
as $f(\bm{A})$ leads us to express $[f^n(\bm{A})]_A$ as a summation/integration 
with respect to $n$ and $nm$ replica variables $\{\bm{c}^a\}$
and $\{\bm{c}^a\circ \bm{u}^{a\sigma} \}$
($a\in \{1,2,\ldots,n\}, \sigma \in\{1,2,\ldots,m\}$), 
which can be evaluated by the saddle point method for $N \to \infty$. 

\begin{figure}
\begin{center}
\includegraphics[width=2.95in]{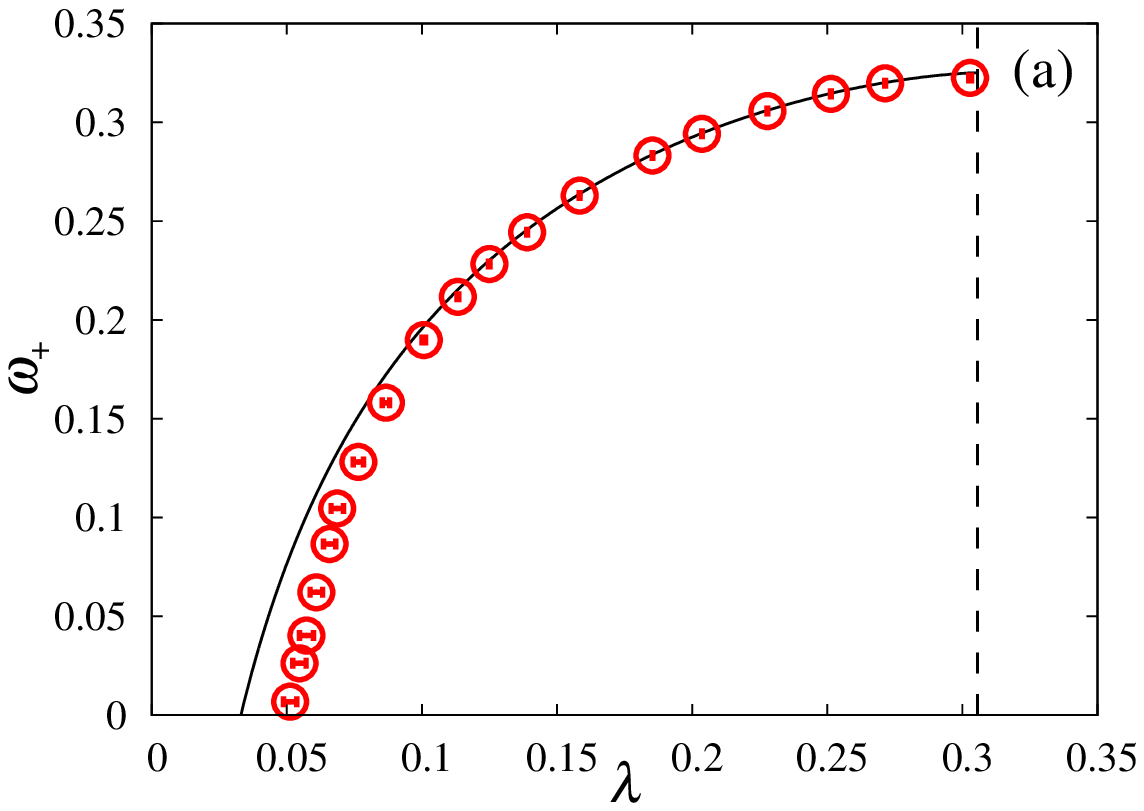}
\includegraphics[width=2.95in]{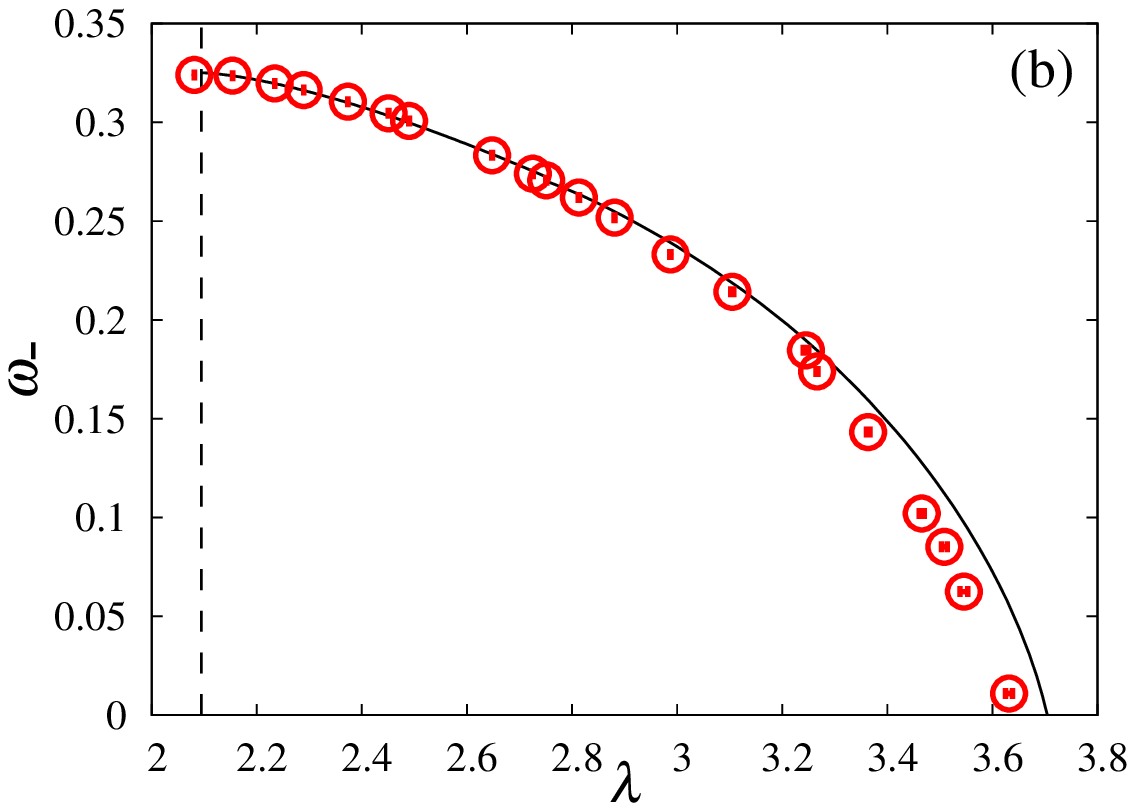}
\end{center}
\caption{Entropy curve for $\alpha=0.5$ and $\rho=0.1$
with (a) $\mu > 0$ and (b) $\mu < 0$.
Circles denote EMC method results.
Vertical lines represent (a) minimum and (b) maximum eigenvalues of MP distribution.}
\label{fig:complexity_alpha05rho01}
\end{figure}

Under the replica symmetric (RS) assumption,
in which the dominant saddle point is assumed to be
invariant against any permutation of the replica indices
$a$ and  $\sigma$ within each of their sets 
$\{1,2,\ldots, n\}$ and $\{1,2,\ldots,m\}$, respectively, 
the resulting functional form of $N^{-1} \log [f^n(\bm{A})]_A$ becomes extendable for 
non-integer $n$ and $m$. Therefore, we insert the expression into 
\Eref{free_entropy} employing the formula of
\Eref{eq:def_replica}, which finally yields
\begin{align}
\nonumber
&\phi(\mu;\rho)=-\frac{\alpha}{2}\log\{\alpha+\chi+\mu(1-q)\}+\frac{\alpha}{2}\log(\alpha+\chi)\\
\nonumber
&-\frac{\alpha\mu q}{2\{\alpha+\chi+\mu(1-q)\}}+\frac{\hat{Q}}{2}-\frac{\hat{q}_1}{2}\Big(1+\frac{\chi}{\mu}\Big)+\frac{\hat{q}_0q}{2}+K\rho\\
&+\!\int \!\!Dz\log\Big\{1+e^{-K}\!\!\int \!\!Dy\exp\!\Big(\frac{(\sqrt{\hat{q}_1\!-\!\hat{q}_0}y\!+\!\sqrt{\hat{q}_0}z)^2}{2\hat{Q}}\Big)\Big\},
\label{eq:RS_free_entropy}
\end{align}
where $\{q,\chi,\hat{Q},\hat{q}_0,\hat{q}_1,K\}$ 
are determined 
to extremize the right hand side, 
and $\int Dz = \int_{-\infty}^{+\infty} \frac{dz}{\sqrt{2\pi}}\exp(-z^2\slash 2)$.
The derivation of \Eref{eq:RS_free_entropy} is shown in 
Appendix \ref{app:replica}.
Entropy densities $\entropy_\pm(\lambda;\rho)$
are derived by applying the inverse Legendre transformation to $\phi(\mu;\rho)$.

\section{Results}

In \Fref{fig:complexity_alpha05rho01},
entropy densities $\entropy_{\pm}$ with
$\alpha=0.5$ and $\rho = 0.1$ are shown
for (a) $\mu < 0$ and (b) $\mu > 0$.
Results of the exchange Monte Carlo (EMC) 
sampling
are represented by circles, and the EMC procedure is summarized in Appendix \ref{app:EMC}.

The values of $\lambda$ when $\mu\to+0$ and $\mu\to-0$,
which are denoted using dashed lines, coincide with
the respective minimum and maximum of the Marchenko-Pastur (MP) distribution's 
support
for the $M\times S$ Gaussian random matrix \cite{MP}.
As the limit of $|\mu| \to 0$ corresponds
to unbiased generation of $M \times S$ Gaussian random matrices, 
the coincidence theoretically supports the adequacy of our analysis.
The slight discrepancy between the theoretical and EMC results in the 
entropy's tails
could be due to the insufficiency of the RS assumption.
The convexity of our entropy suggests
that the RS assumption
exactly creates the entropy curve or extends it outward
\cite{Obuchi2010}.
This is consistent with the EMC method's result,
which indicates that the exact entropy curve is inward when compared to that produced by the RS assumption.
Therefore, the estimated zero-points, $\lambda_{\rm max}^*$ and $\lambda_{\rm min}^*$,
that are provided using the RS assumption,
are meaningful upper and lower bounds, respectively, of the true values.
We call them RS bounds.

\begin{figure}
\begin{center}
\includegraphics[width=3in]{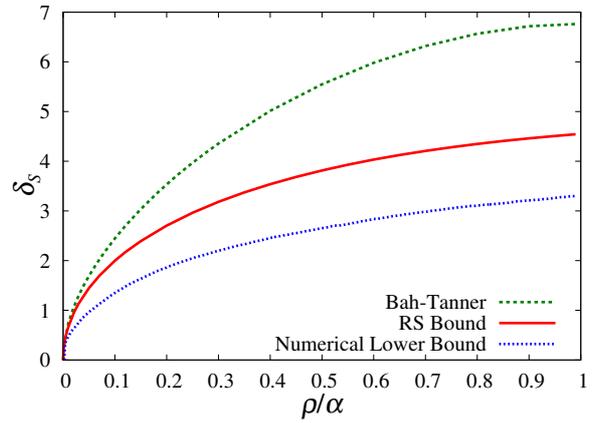}
\end{center}
\caption{Comparison of symmetric RIC for $\alpha=0.5$.
Numerical lower bound is estimated for $N=1000$ and $M=500$.
}
\label{fig:RIP_vs_Bah-Tanner}
\end{figure}

\Fref{fig:RIP_vs_Bah-Tanner}
compares our RS upper bound,
Bah and Tanner's upper bound \cite{CT},
and the RIC numerically obtained lower bound \cite{numericalCS}.
In this example, the symmetric RIC is $\delta_S^{\rm max}$.
Our analysis lowers the upper bound of the RIC, especially
for a large $\rho\slash\alpha$ region.
Over the entire parameter region,
our estimates are consistent with the numerically obtained lower bound.

\begin{figure}
\begin{center}
\includegraphics[width=\figwidth]{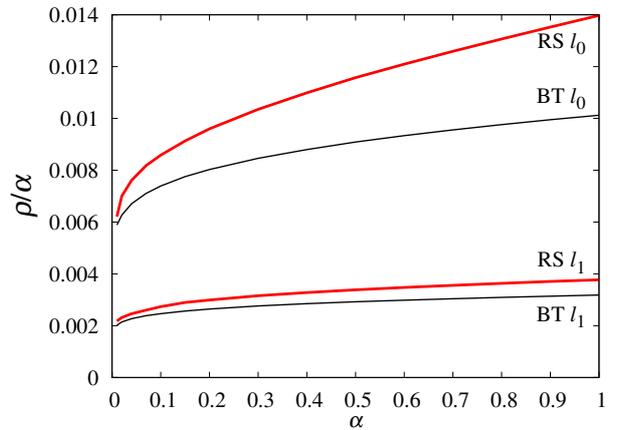}
\end{center}
\caption{$\ell_0$ and $\ell_1$ limits
given by RS RIC.
Black lines represent Bah and Tanner's results, denoted by BT.
}
\label{fig:alpha-rho}
\end{figure}

\Fref{fig:alpha-rho} shows the parameter region
that mathematically supports $\ell_0$ and $\ell_1$ reconstruction according to Theorems 1 and 2.
The region determined by the Bah and Tanner
RIC is indicated using black lines.
The RS bound of the RIC extends
the region in which correct reconstruction is guaranteed,
and further extension may be provided by taking the RSB into account.

\section{Summary and conclusion}

We proposed a theoretical scheme
for the evaluation of restricted isometry constants.
The problem was converted 
to the assessment of entropy density,
and the possible maximum and
minimum eigenvalues, which produce the RIC,
are the entropy's zero-points.
Given a Gaussian random matrix,
we computed the entropy density
using the replica method under the replica symmetric ansatz
and estimated the value of the RIC.
Physically, it has meaning as a bound and
is tighter than existing bounds.
Numerical experiments using the EMC 
sampling
support our analysis.

A more accurate evaluation of the RIC is possible if the RSB is taken into account.
Our scheme is applicable to more general matrices
than Gaussian random matrices as well.

\appendices
\section{RS calculation of free entropy density}
\label{app:replica}

Identities
\begin{align}
1&=\int dq^{(a,\sigma)(b,\tau)}\delta\Big(q^{(a,\sigma)(b,\tau)}-\frac{1}{N}\sum_{i=1}^Nc_i^ac_i^bu_i^{a\sigma}u_i^{b\tau}\Big), 
\label{eq:subshell}
\end{align}
for all combinations of replica indices $(a,\sigma)$ and $(b,\tau)$
$(a,b\!=\!1,2,\ldots,n;\sigma,\tau\!=\!1,2,\ldots,m)$, 
are employed 
in the saddle point assessment of $\phi_\beta(n,m;\rho)\equiv N^{-1}\log [(\sum_{\bm{c}}Z^m(\bm{c},\bm{A};\beta) 
\delta(\sum_{i=1}^N c_i-N\rho))^n]_{A}$.
We assume
that the dominant saddle point is 
of the replica symmetric form as
\begin{align}
q^{(a,\sigma)(b,\tau)}=
\begin{cases}
1 & \mbox{for}~a=b, ~\sigma=\tau \\
q_1 & \mbox{for}~a = b,~\sigma\neq\tau \\
q_0 & \mbox{for}~a\neq b. \\
\end{cases}
\label{eq:RS_assumption}
\end{align}
This means that when $\bm{A}$ is a Gaussian random matrix
of mean 0 and variance $(N\alpha)^{-1}$,
\begin{align}
\nonumber
[s_\mu^{a\sigma}s_\nu^{b\tau}]_A=\alpha^{-1}\delta_{\mu\nu}(\delta_{ab}\delta_{\sigma\tau}+q_1\delta_{ab}(1-\delta_{\sigma\tau})+q_0(1-\delta_{ab}))
\end{align}
holds, where 
$s_\mu^{a\sigma} \equiv \sum_iA_{\mu i}c_i^au_i^{a\sigma}$.
Higher order correlations 
are negligible due to the central limit theorem, 
which indicates that $s_\mu^{a\sigma}$
can be expressed as 
$s_\mu^{a\sigma}=\alpha^{-1\slash 2}(\sqrt{1-q_1}w_\mu^{a\sigma}+\sqrt{q_1-q_0}v_\mu^a+\sqrt{q_0}z_\mu)$,
where $w_\mu^{a\sigma}$, $v_\mu^a$, and $z_\mu$
are i.i.d. Gaussian random variables of zero mean and unit variance.
Replacing $[\cdot]_A$ with average with respect to these Gaussain variables, 
the saddle point evaluation offers an expression of
$\phi_\beta(m,\rho)\equiv \lim_{n\to 0} \frac{\partial}{\partial n} \phi_\beta(n,m;\rho)$, as
\begin{align}
\nonumber
&\phi_\beta(m ;\!\rho)\!=\!\frac{m(\tilde{Q}\!+\!\tilde{q}_1q_1)\!-\!m^2(\tilde{q}_1q_1\!-\!\tilde{q}_0q_0)}{2}\!+\!\!\int\!\! Dz\log\!\Xi_\beta(z)\\
\nonumber
&+\alpha\Big[-\frac{m}{2}\log\!\Big(1\!+\!\frac{\beta(1-q_1)}{\alpha}\Big)\!-\!\frac{1}{2}\log\Big(1\!+\!\frac{\mu(q_1-q_0)}{\alpha\!+\!\beta(1-q_1)}\Big)\\
\label{mid_expression_phi}
&\hspace{1.0cm}-\frac{\mu q_0}{2\{\alpha+\beta(1-q_1)+\mu(q_1-q_0)\}}\Big]+K\rho,
\end{align}
where 
\begin{align}
\nonumber
\Xi_\beta(z)\!=\!1+&\frac{e^{-K}}{(\tilde{Q}+\tilde{q}_1)^{m\slash 2}}
\!\int\!\! Dy\exp\!\Big(\frac{m(\sqrt{\tilde{q}_1\!-\!\tilde{q}_0}y\!+\!\sqrt{\tilde{q}_0}z)^2}{2(\tilde{Q}+\tilde{q}_1)}\Big)
\end{align}
and $\tilde{Q}$, $K$, $\tilde{q}_1$ and $\tilde{q}_0$ are
conjugate variables 
for the integral representations of delta functions
in \Eref{eq:Z_c_def}, \Eref{free_entropy} and \Eref{eq:subshell},
respectively. 
Eq. (\ref{mid_expression_phi}) yields the free entropy density 
as $\phi(\mu;\rho)=\lim_{\beta\to\infty}\phi_\beta(\mu/\beta,\rho)$, 
in which
the variables scale 
so that $\hat{Q} \equiv m(\tilde{Q}+\tilde{q}_1)$,
$\hat{q}_1 \equiv m^2\tilde{q}_1$, $\hat{q}_0 \equiv m^2\tilde{q}_0$, and
$\chi \equiv \beta(1-q_1)$ become $O(1)$. 
This gives the expression of \Eref{eq:RS_free_entropy}.

The variables $\{\chi, q_0,\hat{Q},\hat{q}_1,\hat{q}_0,K\}$ are determined by
extremization 
conditions of the free entropy density \Eref{eq:RS_free_entropy},
\begin{align}
\chi&=\frac{\mu\rho}{\hat{Q}}\\
q&=\int Dz\Big\{\frac{\Xi(z)-1}{\Xi(z)}\frac{\sqrt{\hat{q}_0}z}{\hat{Q}-\hat{\Delta}}\Big\}^2\\
1&=\int Dz\frac{\Xi(z)-1}{\Xi(z)}\Big(\frac{\hat{\Delta}}{\hat{Q}(\hat{Q}-\hat{\Delta})}+\frac{\hat{q}_0z^2}{(\hat{Q}-\hat{\Delta})^2}\Big)\\
\rho&=\int Dz\frac{\Xi(z)-1}{\Xi(z)}\\
\hat{\Delta}&=\frac{\alpha\mu^2(1-q)}{(\alpha+\chi)\{\alpha+\chi+\mu(1-q)\}}\\
\hat{q}_0&=\frac{\alpha\mu^2q}{\{\alpha+\chi+\mu(1-q)\}^2}
\end{align}
where $q =q_0$, $\hat{\Delta}=\hat{q}_1-\hat{q}_0$,
and
$\Xi(z)=\lim_{\beta\to\infty}\Xi_\beta(z)$.

\section{Monte Carlo 
sampling for RIC estimation}
\label{app:EMC}

We employ the exchange Monte Carlo (EMC) 
sampling \cite{EMC}
in order to numerically compute the free entropy density $\phi(\mu|\bm{A};\rho)$
and obtain the entropy density
avoiding the trap of metastable states.
In the EMC approach, 
we prepare $k$ systems, which have the same configuration of $\bm{A}$,
and assign configuration $\bm{c}_i\in\bm{c}_S$
and parameter $\mu_i$
to each system $i=1,\cdots,k$.
The signs of $\{\mu_i\}$ are set to be the same.
Each step of the EMC process updates $\bm{c}_i$
within each system, 
and attempts exchanges between configurations $\bm{c}_i$
and $\bm{c}_{i+1}$.
The probability of transition
from $\bm{c}_i$ to $\bm{c}_i^\prime$ is
given by
$w(\bm{c}_i,\bm{c}_i^\prime)=\min\{\exp(\mu_iN\Delta_{i}),1\},$
where $\Delta_i=\Lambda_{{\rm sgn}(\mu_i)}(\bm{c}_i|\bm{A})-\Lambda_{{\rm sgn}(\mu_i)}(\bm{c}_i^\prime|\bm{A})$.
The probability of an exchange between
systems $\bm{c}_i$ and $\bm{c}_{i+1}$
is given by
$w_{\rm exc}(\bm{c}_i,\bm{c}_{i+1})
=\min\!\{\exp(N(\mu_i\!-\!\mu_{i+1})\Delta_{i,i+1}),1\},$
in which $\Delta_{i,i+1}\!=\!\Lambda_{{\rm sgn}(\mu_i)}(\bm{c}_i|\bm{A})\!-\!\Lambda_{{\rm sgn}(\mu_{i+1})}(\bm{c}_{i+1}|\bm{A})$.
After sufficient updates,
the entire $k$-system is expected to converge to equilibrium distribution
$P_{\rm
 tot}(\{\bm{c}_i\}|\bm{A})\propto\prod_{i=1}^k\exp\{-N\mu_i\Lambda_
 {{\rm sgn}(\mu_i)}(\bm{c}_i|\bm{A})\}$.

The density of states
$W_\pm(\lambda|\bm{A};\rho)\!=\!\sum_{\bm{c}\in\bm{c}_S}\delta(\lambda\slash 2-\Lambda_\pm(\bm{c}|\bm{A}))$
is obtained by applying the
multihistogram method \cite{FS}
using histgrams of $\Lambda_\pm$
obtained by EMC sampling. 
Finally, the free entropy density is calculated:
\begin{align}
\phi(\mu|\bm{A};\rho)=\int d\lambda W_{{\rm sgn}(\mu)}(\lambda|\bm{A};\rho)\exp(-N\mu\lambda\slash2),
\label{eq:app_free_entropy}
\end{align}
and the entropy density is derived by applying the inverse Legendre 
transformation to \Eref{eq:app_free_entropy}.

%

\section*{Acknowledgment}
The authors would like to thank Tomoyuki Obuchi for his helpful comments and
discussions.
This work was partially supported by the
RIKEN SPDR fellowship and by
KAKENHI No. 26880028 (AS),
and KAKENHI No. 25120013 (YK).

\end{document}